**A spatial envelope soliton model of the electron
—propagating within a fictitious waveguide**

Roald Ekholdt[i]

**ABSTRACT:**
The paper proposes an envelope soliton model of the electron, which propagates within a fictitious waveguide. The width of the waveguide varies with the potential. The corresponding non-linear equation is the original Schrödinger equation with the addition of a non-linear term, which is the negative of Bohm's Quantum potential—as it is realised that this potential represents the dispersion ingredient in the linear Schrödinger equation. Thus, the non-linearity cancels the dispersion. This procedure also clarifies the statistical role of the linear equation. The model is based on de Broglie's original electron wave-particle theories. Schrödinger's wave function is revealed as the envelope of the de Broglie wave function.
Since the electron model is based directly on the Special theory of relativity, it may also be used as an illustration of that theory. (In an appendix a somewhat similar model for the photon, based on a virtual circular dielectric rod concept, is suggested—as a very speculative possibility, where the General theory of relativity may play a role.)

**Index**

**1. Introduction**

In this paper an envelope soliton is proposed as a wave-particle model of the electron. The soliton, which is of the spatial breather type, is confined within a fictitious waveguide. While the width in free space is determined by $mc^2$, the width along the waveguide varies with the potential. The particle proper, embodying the charge and spin attributes, zigzag within the envelopes—see figure 1.1. The fictitious waveguide represents the trajectory of the electron. The corresponding equation of motion for the soliton is the same as for classical particles as long as there are no abrupt variations in the potential. In the case of a potential barrier, for instance, the phase of the zigzag motion of the particle will determine whether the particle, and the soliton, is reflected or not. As this phase is not amenable to measurement, at least with

---

[i] Retired. Earlier: Norwegian Defence Research Establishment and Norwegian Telecom Admin. Research Institute, etc. Address: Granstubben 2, N-2020 Skedsmokorset, Norway
 E-mail: rekholdt (at)online. no

existing technology, it qualifies for the term "hidden". This variable causes the statistical character of quantum mechanics.

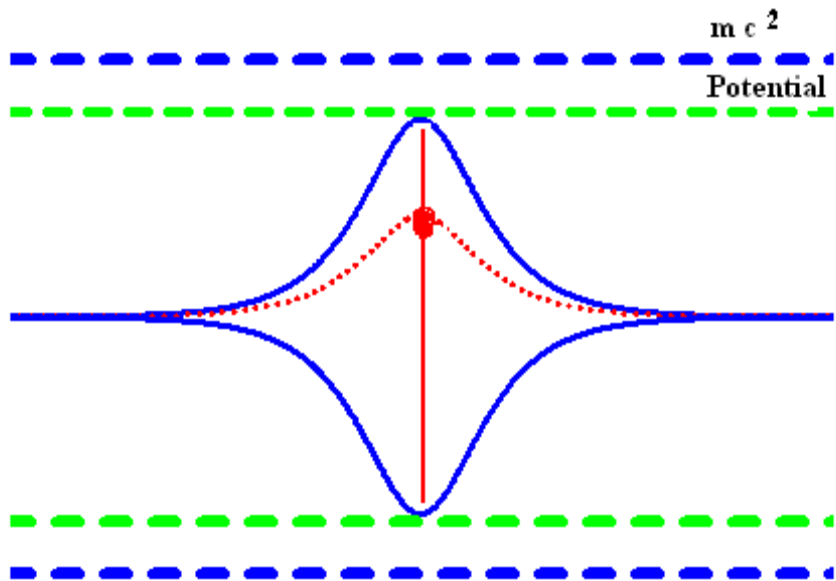

Figure 1.1 The electron as a spatial envelope soliton. The particle and the wave oscillate within the envelopes, which are determined by the rest energy and the potential.

The envelope soliton is governed by a non-linear equation which is the ordinary Schrödinger equation with the addition of a non-linear term. Since the quantum potential of the de Broglie/Bohm theory represents the dispersion of the ordinary Schrödinger equation, and the non-linear term must cancel the dispersion, its negative value is chosen as the non-linear term. This choice was made after efforts to apply a quadratic non-linearity—which is normally used in the so-called Non-linear Schrödinger equation in non-linear theory—failed. The non-linear version of the Schrödinger helps clarify the interpretation the ordinary linear equation.

The waveguide concept follows directly from de Broglie's original proposal, in his doctoral thesis, of the wave attributes of quantum mechanics, while the soliton concept relates to his Pilot wave and Double solution ideas.[1] Unfortunately, these ideas did not come to fruition because they arrived before the microwave waveguide was invented in the nineteen-thirties and before the envelope soliton theory was developed in the seventies.

The waveguide-envelope soliton concept seems to answer fundamental problems regarding the various current interpretations of quantum mechanics:

1. The indeterminism.
2. The collapse of the wave-function concept is eliminated, since the electron is confined to the waveguide.
3. The Bohr-Sommerfeld atomic theory may be revived—with the waveguide replacing the ideal path.
4. The hidden variable idea is revitalized.
5. The probability regarding wave reflections et cetera.
6. The Heisenberg uncertainty relations seem to be analogue to the ambiguity relation for pulse-doppler radars.

7. The Quantum Mechanics and the Special theory of relativity are reconciled, since the waveguide model is directly based on the latter.

Although this model is local, the problems regarding spin and the polarisation of photons remain. One may hope that the electron model may be developed to include spin, and that the waveguide concept may be applied to photons as well. In case one may possibly get a better understanding of the EPR phenomena. For good measure the polarisation of photons is touched upon, lightly and very speculatively, in appendix B.

The paper is written in the spirit of Einstein's consoling words to Louis de Broglie in Brussels in 1927 after the Solvay congress, where he was ridiculed by Wolfgang Pauli: *"Continue! It's you who are on the right course. […] a physical theory should, except for all calculations, be possible to illustrate by pictures so simple that a child should be able to understand them."*[2] But de Broglie experienced Walter Bagehot's words:*"…one of the greatest pains to human nature is the pain of a new idea."* And de Broglie gave up his project, until his quantum potential idea was revived, independently, by David Bohm in 1952.[3] The following quotation from Bell inspired the soliton proposal, which matches his prediction remarkably well:[4] *"The pragmatic attitude [following the Copenhagen interpretation], because of its great success and immense continuing fruitfulness, must be held in high respect. Moreover it seems to me that in the course of time one may find that because of technical pragmatic progress the 'Problem of Interpretation of Quantum Mechanics' has been encircled. And the solution, invisible from the front, may be seen from the back. For the present, the problem is there, and some of us will not be able to resist paying attention to it. The nonlinear Schrödinger equation seems to me to be the best hope for a precisely formulated theory which is very close to the pragmatic version."*

**2. De Broglie's basic theory**

De Broglie took as his point of departure that quantum mechanics was based on the Planck-Einstein radiation quant, later called the photon, and that the Special theory of relativity played a decisive role.[5] Further, he believed that the electron and the photon were closely related, and that both had a wave-particle nature. He thus assumed that the electron's rest energy could be represented as $E_o = m_o c^2 = hf_o$ and the momentum as $P = h/\lambda$ — where $f_o$ and $\lambda$ represent frequency and wavelength.

But while the photon propagates with the velocity of light, the electron may even be at rest. To solve this problem de Broglie introduced the hypothesis that the electron, considered as a particle, had an inner vibration with a frequency that varied as the relativistic clock frequency

$$f_{clock} = f_o \sqrt{1-(v/c)^2} \qquad (2.1)$$

where *v* is the electron's velocity.

Moreover, he assumed that a moving electron's energy varied relativistically as $E(v) = hf$, and thus that the frequency of the wave varied as

$$f_{wave} = f_o / \sqrt{1-(v/c)^2} \ . \qquad (2.2)$$

He thought that this wave propagated all over physical space, and even at higher velocities than the speed of light. But in his opinion this was not in conflict with the theory of relativity since the wave did not carry energy. This strange type of wave has quantum theory struggled with all the time after Schrödinger removed the particle and only this type of waves remained.

Unfortunately, in view of his later insight into microwave waveguides, de Broglie could not at the time realize the waveguide analogy. During the First World War he served at a military radio telegraph station in the Eiffel Tower, and his co-worker and biographer Georges Lochak writes: *"From the war years he had a constant interest in the applications of physics, in particular the development within radio technology, which he followed closely, to the latest progress of waveguides..."*—see reference 1. Actually, Feynman mentions this analogy briefly in the chapter on waveguide in his lecture series—but he does not seem to have elaborated on it. [6]

**3. The waveguide analogy**

Figure 3.1 depicts a topside view of a waveguide, similar to the common microwave rectangular metallic type used for the $TE_{10}$- mode, where we hypothetically assume an electron, consisting of a particle and a piece of a wave-front, zigzags between the side-walls at the speed of light *c*.

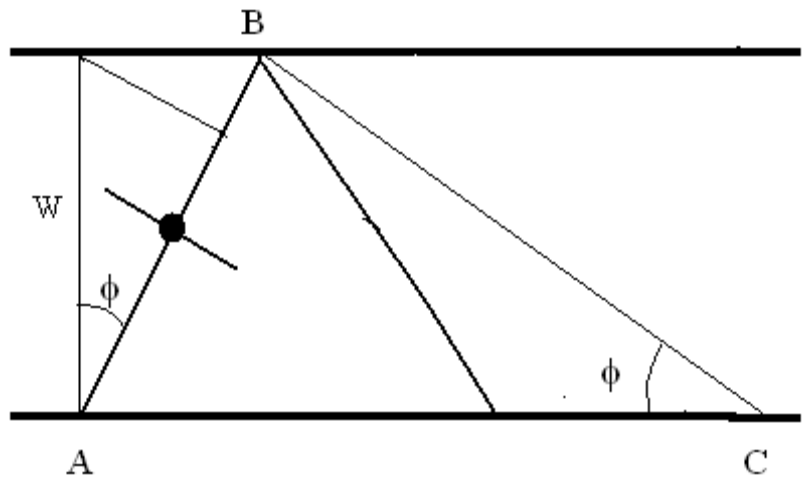

Figure 3.1. The particle with a wavefront zigzaging between the waveguide's sidewalls. (Interestingly, this figure resembles Penrose's zigzag picture of Dirac's electron theory.[7] )

The relation between the electron's velocity along the waveguide, $v$, and the angle $\phi$ is

$$v = c \sin \phi = c \sqrt{1 - \left(\frac{c}{2fw}\right)^2} . \quad (3.1)$$

The zigzag period is thus

$$t_{zigzag} = \frac{1}{f_{zigzag}} = \frac{\cos \phi}{f_o} = \frac{1}{f_o}\sqrt{1-(v/c)^2} \quad (3.2)$$

i.e.

$$f_{zigzag} = f_{clock} = f_o \sqrt{1-(v/c)^2} \quad (3.3)$$

while the length of the zigzag period along the waveguide is $L_{zigzag} = 2w\tan\phi$.

The wavelength of a continuous, monochromatic signal along the waveguide would have been

$$\lambda = 2w\cos\phi = 2w\gamma = T_o c\gamma \quad (3.4)$$

$$\text{where } \gamma = \sqrt{1-(v/c)^2}$$

Here $v$ represents the effective propagation velocity of the signal's energy. The frequency is thus

$$f_{wave} = \frac{c}{2w\gamma} = f_o/\gamma \quad (3.5)$$

which is equal to equation (2.2).

From the triangle ABC in figure 3.1 we see that the phase velocity of the wave along the waveguide is

$$V_{phase} = c/\sin\phi. \quad (3.6)$$

Thus

$$vV_{phase} = c^2. \quad (3.7)$$

The cut-off frequency $f_o = m_o c^2/h$ represents the lowest possible frequency for wave propagation, except for evanescent waves over short distances—equivalent to tunnelling in quantum mechanics. The waveguide dimension is thus

$$w = \frac{c}{2f_o} = \frac{h}{2m_o c} = 1.21 \cdot 10^{-11} \text{ m} \quad (3.8)$$

The rest mass thus determines the width of the waveguide—in the zero potential case.

## 4. Similarity with Bohr's orbits in the hydrogen atom

Since the details of de Broglie's development of Bohr's empirical quantization rule for the angular momentum in the hydrogen case largely appears to have been forgotten, it seems valuable to illustrate the waveguide analogy on that simple case—following Lochak's rendition; see appendix I. There we replace Bohr's ideal paths, with no transverse dimension, by waveguides. Without making extra obscure assumptions as Bohr and de Broglie had to do, we arrive at Bohr's non-relativistic empirically based angular momentum relation.

In conjunction with Bohr's model of the hydrogen atom we note the analogy regarding the energy relation between the energy levels and the energy of corresponding photons and the frequency relations of frequency converters. *This indicates the likelihood of a non-linearity effect in the atom.*

**5. Wave equations**

De Broglie might have established a wave equation directly by using the relativistic energy relation $E^2 = (m_o c^2)^2 + (cP)^2 = (hf_o)^2 + (c h/\lambda)^2$, where $P$ represents the linear momentum, and further the corresponding dispersion relation

$$f^2 = f_o^2 + (ck)^2 . \tag{5.1}$$

see figure 3.1.

The corresponding wave equation is

$$\frac{\partial^2 \psi(z,t)}{\partial t^2} + \frac{\partial^2 \psi(z,t)}{\partial z^2} + f_o^2 \psi(z,t) = 0 . \tag{5.2}$$

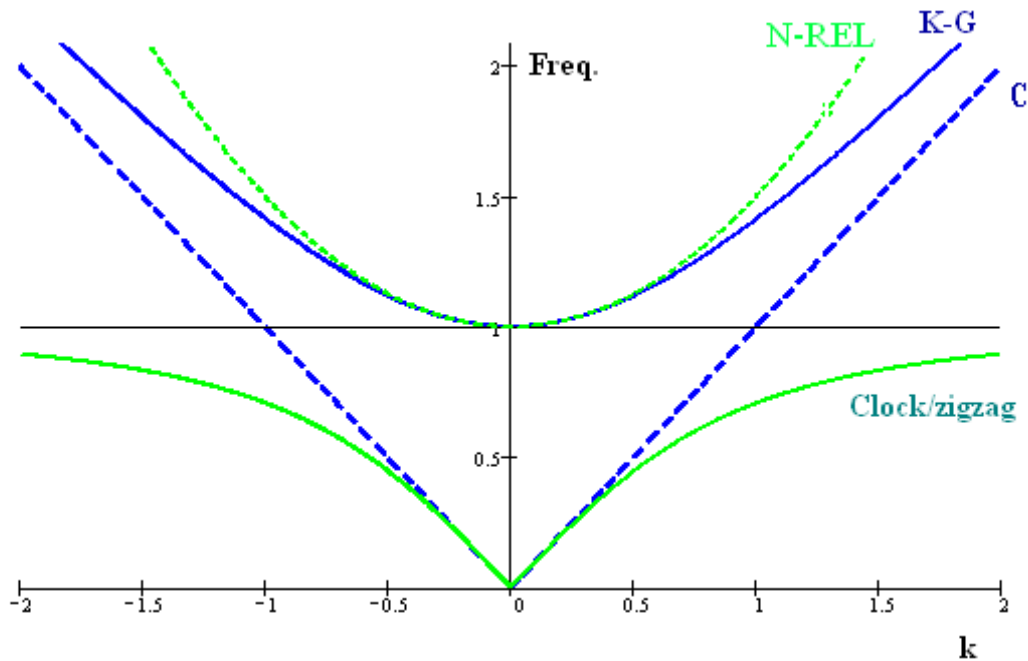

Figure 5.1. Dispersion relations normalised to $f_o = 1$ and $c=1$. Note that the Schrödinger case here has the rest energy included—represented by $f_o = 1$, but no frequency shift due to a potential energy.

But before de Broglie, in parallel with others, in 1926 developed this so-called Klein-Gordon equation, Schrödinger in 1925 had successfully presented his non-relativistic equations.[8] He probably first tried the relativistic equation, but could not find agreement with empiric data, and did not publish this result—and the notes on this work disappeared when he escaped from Austria to Dublin in 1939. By a rather *ad hoc* process he developed his non-relativistic equation

$$i\hbar \frac{\partial \psi}{\partial t} + \frac{\hbar^2}{2m} \frac{\partial^2 \psi}{\partial z^2} - V(z)\psi = 0 \tag{5.3}$$

where $\hbar = h/2\pi$ and $V(z)$ represents the potential. The parameter $i = \sqrt{-1}$ makes his equation very strange.

From our point of view this equation could have been established by introducing an approximation to the dispersion relation for low velocities, i.e. low energies, and by including the potential

$$\omega = \omega_o + V(z)/\hbar + \frac{(ck)^2}{2\hbar\omega_o}. \tag{5.4}$$

But Schrödinger implicitly kept the wave function as $\psi = e^{i(\omega \cdot t - kz)}$, which introduces $i = \sqrt{-1}$ into his wave equation. Equation (5.4) corresponds to the classical energy equation—except for the rest energy: $E = E_o + V + m_o v^2/2$. See figure 5.1. The velocity of an electron viewed as a pulse or a wave packet is then

$$v_{group} = \frac{\partial \omega}{\partial k} = c^2 k / \omega_o \tag{5.5}$$

Clearly, our waveguide model may be used to determine the bending of the ray due to a transversal variation of the potential.

As seen in equation (5.3), in addition to consider non-relativistic velocities only, he also omitted the rest energy—which certainly was natural to do from a classical point of view. But thus he inadvertently "blew" away the sidewalls of the waveguide. Consequently, Schrödinger opened the whole universe for the wave function, which in the waveguide model is confined inside the guide. By doing that he destroyed the analogy with Bohr's orbital model of the hydrogen atom, and thus created great problems regarding how to understand quantum mechanics. *The most serious consequence was, however, that it required the problematic artificial concept of the collapse of the wave function to relate the wave-function results of his equation to that of measurements.*

Further, by removing the carrier in the wave packet the Schrödinger equation only represents the amplitude envelope of the pulse. Indeed, Schrödinger considered the pulse idea, but Lorentz dissuaded him from it by arguing that the pulse would undergo dispersion. Of course they did not know the soliton concept, where nonlinearity fully compensates for the dispersion. It is ironic that for the so-called envelope solitons—which are applied in optical fibres communication, the amplitude is described by a non-linear version the Schrödinger equation.

## 6. Wave packet as an envelope soliton

In considering the wave-particle model we have so far more or less tacitly assumed a particle and, in principle, an unlimited continuous wave. In the nineteen fifties de Broglie in his double solution introduced non-linearity, but this was ten years before the great development of non-linear theory took off—especially by the introduction of the soliton concept. Early in the seventies the theory of the envelope soliton was developed. Since we are dealing with a waveguide, an envelope solition solution seems an interesting idea for the wave part of the wave-particle concept.

The envelope soliton is normally associated with the so-called non-linear Schrödinger (NLS) equation[9], here in normalized form,

$$i\frac{\partial \varphi}{\partial t} + \frac{\partial^2 \varphi}{\partial z^2} + 2|\varphi|^2 \varphi = 0 \qquad (6.1)$$

The exact solution of this non-linear equation is

$$\varphi(z,t) = a \cdot \exp[ivx/2 + i(a^2 - v^2/4)t] \cdot sech[a(z - vt - z_o)] \qquad (6.2)$$

which is a so-called breather moving at velocity $v$. For $v = 0$ the solition is reduced to a so-called stationary breather

$$\varphi(z,t) = a \cdot \exp(ia^2 t) \cdot sech[a(z)] \qquad (6.3)$$

But the quadratic non-linearity seems to make the NLS equation incompatible with the normal Schrödinger equation. Bohm's quantum potential procedure, however, seems to point to a solution to this problem.[10] (In 1927 de Broglie proposed the same procedure.)

Wishing to consider the classical limit in quantum theory, Bohm assumed the WBK approximation, and introduced the wave function as

$$\psi = R \exp(iS/\hbar) \qquad (6.4)$$

where $R$ and $S$ are two real functions. By substitution into equation (5.3), and by separation of the real and imaginary parts the two following equations result:

$$\frac{\partial S}{\partial t} + \frac{1}{2m}(\frac{\partial S}{\partial z})^2 + V - \frac{\hbar^2}{2m}(\frac{\partial^2 R}{\partial z^2})/R = 0 \qquad (6.5)$$

$$\frac{\partial R}{\partial t} + \frac{1}{2m} \cdot [R^2(\frac{\partial^2 S}{\partial z^2}) - 2\frac{\partial R}{\partial z}\frac{\partial S}{\partial z}] = 0 \qquad (6.6)$$

Except for the last term and by noting that *S* has the dimension of action, equation (6.5) is equal to the Hamilton-Jacobi equation of classical mechanics, which represents a particle with momentum $p = \frac{\partial S}{\partial z}$. Bohm called the last term a quantum potential *Q*

$$Q = -\frac{\hbar^2}{2m}(\frac{\partial^2 R}{\partial z^2})/R \tag{6.7}$$

which he interpreted as containing the difference between quantum and classical mechanics. *Since he maintained the normal quantum mechanics statistical interpretation of the wave-function, Q represented for him the instantaneous so-called entanglement over all space.*

Since the quantum potential derives from the term in the Schrödinger equation that represents dispersion, and we seek a non-linear term that cancels it, we heuristically tries to add its negative into the ordinary equation:

$$i\hbar\frac{\partial \psi}{\partial t} + \frac{\hbar^2}{2m}\frac{\partial^2 \psi}{\partial z^2} - V(z)\psi + [\frac{\hbar^2}{2m}\left(\frac{\partial^2 R}{\partial z^2}\right)/R]\psi = 0 \tag{6.8}$$

Thus *Q* is eliminated from equation (6.5), which becomes equal to the classical Hamilton-Jacobi equation

$$\frac{\partial S}{\partial t} + \frac{1}{2m}(\frac{\partial S}{\partial z})^2 + V = 0 \tag{6.9}$$

Hence, the equation of motion for the soliton is the same as for a classical particle.

Referring to NLS, we choose the envelope as

$$R = r \cdot sech[a(z - v_e t)] \tag{6.10}$$

This results in the non-linear equation

$$i\hbar\frac{\partial \psi}{\partial t} + \frac{\hbar^2}{2m}\frac{\partial^2 \psi}{\partial z^2} - V\psi + \frac{\hbar^2 r a^{-2}}{2m}\{tanh^2[a(z - v_e t)] - 2sech^2[a(z - v_e t)]\}\psi = 0 \tag{6.11}$$

To get the correct value of the non-linear term we choose *a=1*.

We may interpret the *R(z,t)* soliton as a spatial breather, where the particle follows the wave oscillating transversally at the zigzag frequency. Charge and spin are attributes of the particle. See figure 6.1. The amplitude is determined by *mc²* and *V*:

$$r = \frac{ch}{4(m_o c^2 + V)} \tag{6.12}$$

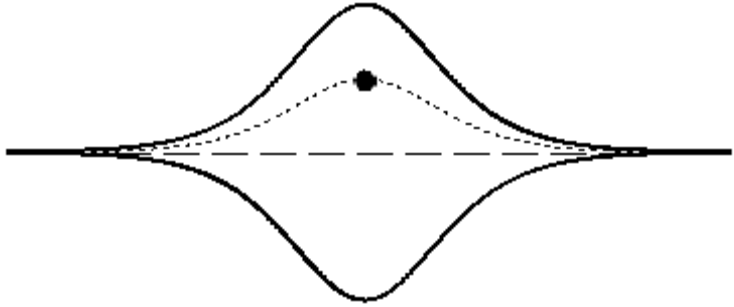

Figure 6.1. The soliton may be regarded as a spatial breather, where the particle follows the wave oscillating transversally at the zigzag frequency.

**7. Interpretation of the difference between our non-linear version of the Schrödinger equation and the original linear equation**

The developed non-linear equation represents the equation of motion for an envelope soliton, which may be seen as belonging to physical reality—a ***beable*** in Bell's terminology. The equation may accordingly be used to determine the probability that a particle *is* at a particular point along its trajectory.

The original Schrödinger equation may, as usual, be considered to represent a statistical ensemble of fictitious, non-interacting point particles—which makes available the probability of *finding* a fictitious particle at a particular point. The ensemble may be seen as filling up our wave-guide model, and thus the continuous wave concept replace the single envelope soliton concept. The density of the fictitious point particles depends on their velocity—which, as we have seen earlier, depends on the potential. The role of the equation is here primarily to represent the velocity's relation to the potential—it may thus be considered as a kinematical equation. From our point of view the quantum potential should only be interpreted as a local phenomenon—not as entanglement of the physically fictitious ensemble over all space.

In cases when the potential varies considerably within a wavelength, reflections will occur. The wave-function then gives the statistical relation between the reflected and the continuing electrons. To illustrate how the mechanism that decides what happens to the individual electron, we assume a potential barrier. This means that there is an abrupt reduction of the waveguide width. We have than to consider the zigzag phase of the particle part of the electron, and how it hits the interface with the barrier. If it hits the barrier itself, it is reflected; if it hits the gap left in the guide, it continues. If the continuing waveguide only opens for evanescent propagation there will be reflection, as well as tunnelling if the restriction is relative short. (Evanescent propagation and tunnelling need further study.) As the zigzag phase is not amenable to measurement, at least with existing technology, it qualifies for the term "hidden". In the linear equation case we get superposition of the wave functions of propagation in opposite directions.

## 8. Concluding remarks

The model seems to reduce the width of the gap between classical and quantum mechanics, while explaining the differences. For instance, it revives the original Bohr-Sommerfeld classical atom model by explaining the roles of classical mechanics as well as of wave mechanic in the prescription of the discrete orbital structure. Moreover, it may provide answers to fundamental problems regarding the various current interpretations of quantum mechanics:

1. The collapse of the wavefunction seems to be eliminated, since the electron is confined to the waveguide
2. The indeterminism, as a basic principle is likewise eliminated—even if limited measurement possibilities prevent the resolution of the apparent statistical phenomena into elementary events.
3. The hidden variable idea may be revitalized.
4. The probability regarding wave reflections are clarified.
5. The Heisenberg uncertainty relations seem to be analogue to the Ambiguity relation for pulse-doppler radars[11].
6. The Quantum Mechanics and the Special theory of relativity are reconciled, since the waveguide model is directly based on the latter.

Although this model is local, the problems regarding spin and the polarisation of photons remain. One may hope that the electron model may be developed to include spin, and that the waveguide concept may be applied to photons as well. In case one may possibly get a better understanding of the EPR, e.g. the Aspect experiments.[12]. For good measure, the polarisation of photons in a dielectric cylindrical rod waveguide model is considered, lightly and very speculatively, in appendix B.

**Appendix A**

**Similarity with Bohr's orbits in the hydrogen atom**

Since the details of de Broglie's development of Bohr's empiric quantization rule for the angular momentum in the hydrogen case largely appears to have been forgotten, it seems valuable to illustrate our waveguide analogy on that simple case—following Lochak's rendition; see reference 3.

In his planetary model Bohr applied classical mechanics, where the centrifugal force balances the attractive force of the nucleus:

$$\frac{mv_e^2}{r} = \frac{e^2}{r^2} \quad \text{i.e.} \quad mv_e^2 r = v_e M = e^2 \tag{10}$$

where $M$ represents the angular momentum. The classical energy of the hydrogen atom is thus

$$E = \frac{mv_e^2}{2} - \frac{e^2}{r} = -\frac{mv_e^2}{2} \tag{11}$$

From the empiric spectral lines Bohr established the relation $M = Nh/2\pi$, where $N$ is an integer quantum number.

De Broglie based his treatment of the hydrogen atom on his Accordance of phases law, which he considered as his most important achievement in quantum theory—reference 5, pages 3 - 4. He found that for a Galilean observer the phase of the internal vibration and the phase of the wave along the path were identical at any time:

$$\varphi_{clock}(z) = f_{clock} t = f_{clock} \frac{z}{v_e} \qquad (12)$$

and

$$\varphi_{wave}(z) = f\gamma(t - V_{phase} z/c^2) = \frac{f_o}{\gamma}(\frac{z}{v_e} - \frac{v_e z}{c^2}) = f_o \gamma \frac{z}{v_e} = f_{clock} \frac{z}{v_e} = \varphi_{clock} \qquad (13)$$

where he applies the Lorentz transform for time.

Again it seems unfortunate that he did not know the waveguide analogy, because the accordance of the two phases is evident from the triangle ABC in figure 3.1, since the phase variation between A and B is equal to the wave's phase shift between A and C.

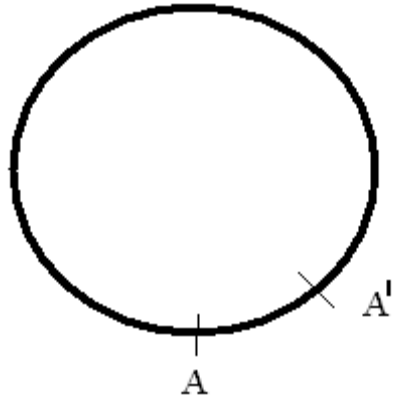

Figure A1.  After the electron has gone a full circle from the origin A it has to move an extra arch A-A' for their joint equal phase at the origin of the wave and the zigzag movement to reappear.

De Broglie chose a point A, see figure A1, as a common origin for the electron and the wave to start at the same time and phase. (The wave is considered as fictitious outside the electron itself.) When, after the time $T$, the particle has made a full circle, it is, however, very unlikely that that the waves, which phase velocity is much higher than the particle velocity, at A has the original common phase. He therefore assumed that the particle had to move an additional arch A - A' in an extra time $\tau$ for this to occur; thus

$$V_{phase} \tau = c^2 \tau / v_e = (\tau + T) v_e \qquad (14)$$

But due to the wave's much higher phase velocity than the particle's, it makes a full extra circle, thus

$$\tau = \frac{v_e^2}{c^2 - v_e^2} T \quad (15)$$

De Broglie then supposed that stationarity required that the phase from A to A' had run through $2\pi N$ radians — where $N$ is an integer, i.e..

$$2\pi \cdot f_{clock} \tau = 2\pi \frac{m_o c^2}{h} \gamma \frac{v_e^2}{c^2 - v_e} = 2\pi N \quad (16)$$

This means that we have $m_o v_e^2 T / \gamma = Nh$ —or nonrelativistically $m_o v_e^2 T = Nh$, which is Bohr's empirically based relation. Now $m_o v_e^2 T = m_o v_e L_{orbit} = h L_{orbit} / \lambda = Nh$, and thus $L_{orbit} = N\lambda$. That the wave pattern moves $\tau \cdot v_e$ per round trip has little importance for a circle path, but for Sommerfeld's elliptic orbits it influences the precession movement of the ellipses' main axis.

.
**Appendix B**

**Investigating the possibility of the photon trajectory as a cylindrical dielectric rod, or fibre, waveguide**

**1. Introduction**

Our interpretation of the trajectory of the electron based on an analogy with a common microwave waveguide makes it tempting to look for a similar wave-particle model for the photon. The following is a brief speculative sketch. Their differences regarding *inter alia* mass, charge, velocities and frequencies amounts to serious obstacles:

1. No known mass.
2. No charge.
3. Velocity *c* in free space.
4. Frequencies down to zero

The low frequency wave problem may possibly indicate the lowest mode of a circular dielectric rod or fibre waveguide—the N1 mode, as this mode has zero as the cut-off frequency.[13] At low frequencies this mode has an evanescent part that reaches extends far outside the mode—see figure B1. This feature may possibly explain the dual slit experiment. The particle trajectory, which may be considered to be confined to the inside of the rod, passes thru one of the slits. The particle may hence be seen as a soliton belonging to the next mode, the N2, which have a higher cut off frequency, and have dispersion diagram similar to the $TE_{10}$ – mode of the electron model. These modes have sinusoidal variations of the fields about the circumference, which may be interpreted as due to contra-rotating fields; this may explain the spin *h*.

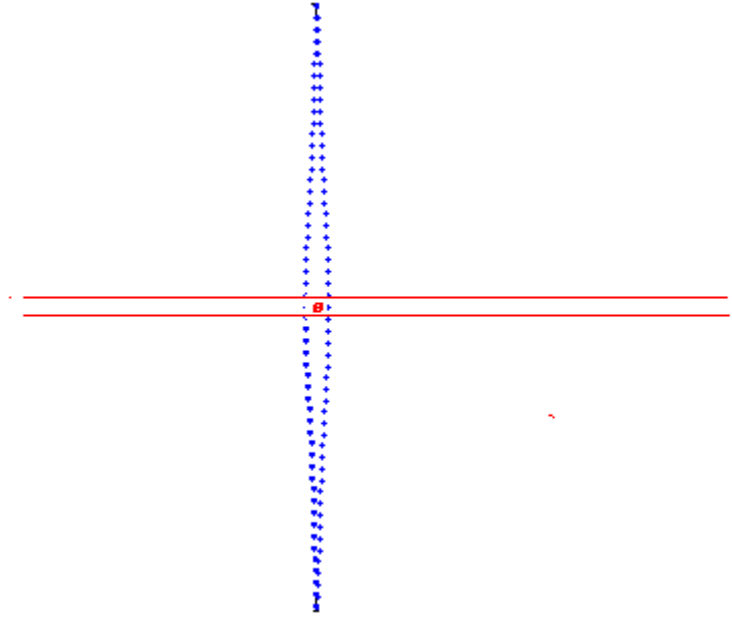

Figure B1. Illustration of a circular dielectric rod, or fibre, waveguide and evanescent part of the N1 mode.

The confinement is considered to be brought about by total reflection, which requires high frequencies and group velocities very near to *c*. Moreover, the N1 mode may be viewed as coupled to the zigzag frequencies at low frequencies. The frequencies *f* in the N2 mode are related as

$$f_{zigzag} = f_o^2 / f \qquad (B.1)$$

and thus

$$hf_{photon} = hf_{zigzag} = hf_o^2 / f \qquad (B.2)$$

Now the question of what determines the diameter of the rod and the cut-off frequency $f_o$ arises. Since only gravity, according to Einstein, influences light in otherwise free space, it seems necessary to go to the General theory of relativity here—instead of the Special theory of relative in the electron case. But that is outside the scope of this paper.

The fields of the N1 mode are of the vector type, and thus the polarisation of the evanescent field arising from the coupling with the soliton, corresponds to, and behaves like the polarisation of continuous electromagnetic waves. The Malus rule is thus valid. With such a model of the photon the Aspect type EPR experiments might perhaps turn out to be due to a common source or a common cause phenomenon.

References:

[1] L. de Broglie, *Interpretation of quantum mechanics by the double solution theory,* Annales de la Fondation Louis de Broglie, Vol. 12, no. 4, 1987

[2] G.Lochak, *Louis de Broglie – Un prince de la science*, Flammarion, Paris, p141, 1992

[3] D.Bohm and B.J.Hiley, *The Undivided Universe,* Routledge, London, 1993, pp 134 - 159

[4] J.S.Bell: *Speakable and unspeakable in quantum mechanics,* Cambridge University Press, 1987, p 194

[5] G.Lochak, *De Broglie's Initial Conceptions of de Broglie Waves*, in S.Diner, D.Fargue, G.Lochak, F.Selleri, (Editors), *The Wave-Particle Dualism,* Kluwer Academic Publishers, 1983, pp 1-8—and further references therein.

[6] R.P.Feynman et al, *Feynman Lectures on Physics,* Addison-Wesley, 1964, Boston, MA, Chapter 24, p 7.

[7] R.Penrose, *The road to reality,* Jonathan Cape, London, 2004, pp 628 - 32

[8] W.Moore, *A life of Erwin Schrödinger,* Cambridge University Press, 1994

[9] A. Scott, *Nonlinear Science,* Oxford University Press, Oxford, 1999, pp 94 – 100

[10] D. Bohm and B.J. Hiley, *The undivided universe, Routledge, London, 1993, pp28 – 30.*

[11] Merrill I. Skolnik, *Radar Handbook,* McGraw-Hill, New York

[12] A.Aspect, J.Dalibard, G.Roger, *Experimental Tests of Bell's Inequality Using Time-Varying Analyzers,* Physical Review Letters 49; pp 1804 – 1807.

[13] See e.g. S.Ramo & J.R.Whinnery, *Fields and Waves in Modern Radio*, Wiley, New York, 1953, pp 388-93. Or in newer editions: S.Ramo, J.R.Whinnery, and T.Van Duzer, *Fields and Waves in Communications Electronics,* Wiley.